\documentclass[11pt,fleqn]{article}
\usepackage{bbm}
\usepackage[numbers,sort&compress]{natbib}
\usepackage{graphicx}
\usepackage{amsmath}
\usepackage{amssymb}
\usepackage{indentfirst}
\usepackage{bbm}
\usepackage{mathrsfs}
\newcommand{\Tr}{\ensuremath{\mathop{\mathrm{Tr}}}}

\def\ni{\noindent}
\def\be{\begin{equation}}
\def\ee{\end{equation}}
\def\bea{\begin{eqnarray}}
\def\eea{\end{eqnarray}}
\def\bsp{\be\begin{split}}
\def\la{\langle}
\def\ra{\rangle}

\def\bes{\be  \begin{split}}

\def\G{\Gamma}

\def\k{\kappa}
\def\d{\delta}
\def\e{\epsilon}
\def\m{\mu}

\def\n{\nu}

\def\l{\lambda}
\def\t{\tau}

\def\w{\wedge}
\def\vp{\varphi}

\def\z{zigzag }
\def\w{Wilson loop }
\def\th{\theta}

\def \nz{{N_z}}
\def \nz{{N}}
\makeatletter

\newcommand{\Rmnum}[1]{\expandafter\@slowromancap\romannumeral #1@}
\makeatother

\renewcommand{\title}[1]{\vbox{\center\LARGE{#1}}\vspace{5mm}}
\renewcommand{\author}[1]{\vbox{\center\large{#1}}\vspace{5mm}}
\newcommand{\address}[1]{\vbox{\center\em#1}}
\newcommand{\email}[1]{\vbox{\center\tt#1}\vspace{5mm}}

\allowdisplaybreaks

\begin{document}

\begin{titlepage}
\hfill {\tt HU-EP-09/19}\\
\title{\vspace{1.0in} {\bf Null Zig-Zag Wilson Loops in ${\cal N}=4$ SYM}}

\author{Xie Zhifeng${}^{1,2}$}

\address{${}^1$Institute of Theoretical Physics, \\Chinese Academy of
             sciences, Beijing 1000190, China}

\address{${}^2$Institut f\"ur Physik, Humboldt-Universit\"at zu Berlin,\\ Newtonstr.
12, 12489 Berlin, Germany}

\email{xzhf@itp.ac.cn}

\abstract{\ni In planar ${\cal N}=4$ supersymmetric Yang-Mills theory
  we have studied supersymmetric Wilson loops composed of a large
  number of light-like segments, i.e., null zig-zags. These contours
  oscillate around smooth underlying spacelike paths. At one-loop in
  perturbation theory we have compared the finite part of the
  expectation value of null zig-zags to the finite part of the
  expectation value of non-scalar-coupled Wilson loops whose contours
  are the underlying smooth spacelike paths. In arXiv:0710.1060
  [hep-th] it was argued that these quantities are equal for the case
  of a rectangular Wilson loop. Here we present a modest extension of
  this result to zig-zags of circular shape and zig-zags following
  non-parallel, disconnected line segments and show analytically
  that the one-loop finite part is indeed that given by the smooth
  spacelike Wilson loop without coupling to scalars which the zig-zag
  contour approximates. We make some comments regarding the
  generalization to arbitrary shapes.}

\end{titlepage}
\section{Introduction and Summary}

Wilson loops play a priviledged role in the AdS/CFT correspondence
\cite{Maldacena:1998im}. Their study has branched into a wide
variety of topics in the correspondence including most recently
scattering amplitudes and integrability. Alday and Maldacena put
forward a Wilson loop composed of light-like segments as the strong
coupling dual of planar gluon scattering
amplitudes~\cite{Alday:2007hr,Alday:2007he}. At strong 't Hooft
coupling, such a Wilson loop is described by a macroscopic
fundamental string whose worldsheet is a minimal area embedding in
$AdS_5$ with boundary given by the Wilson loop contour. Amazingly,
when that same Wilson loop was considered at weak coupling in planar
${\cal N}=4$ SYM perturbation theory, it agreed with the gluon
scattering amplitudes. The correspondence between the light-like
Wilson loop and gluon scattering amplitudes appears in fact to be an
all-orders statement in planar ${\cal N}=4$ SYM and its string
theory dual, i.e. IIB strings on $AdS_5 \times S^5$, \cite{wlrefs,
Drummond:2008aq, Drummond:2007aua, Brandhuber:2007yx,
Drummond:2007cf, Berkovits:2008ic}.
\subsection{Introduction}
It is well known that the Maldacena-Wilson loop, given in general by
\be\label{W} W = \frac{1}{N} {\Tr}_R \,{\cal P}\, \exp \oint_C d\t\,
\Bigl(i\,\dot x^\m A_\m + |\dot x| \Theta_I\Phi_I \Bigr) \ee
may be thought of as the holonomy of an infinitely massive $W$-boson
obtained through Higgsing the $SU(N+1)$ theory to $SU(N)$, or
equivalently, geometrically, by separating a single D3-brane from a
stack of $N+1$; the long string stretching back to the stack being
the $W$-boson. Although the AdS/CFT correspondence is in general
better set in Euclidean space, it is interesting to consider
Lorentzian contours. In particular, choosing $x^\m$ null produces a
Wilson loop without coupling to scalar fields
\be\label{Wnull} W_\emptyset = \frac{1}{N} {\Tr}_R \,{\cal P}\, \exp
i\oint_C d\t\, \dot x^\m A_\m \ee
Interpreting such an object as an holonomy of an infinitely massive
particle appears strange, since we are now requiring its worldline
to be null. It turns out however, that another interpretation of
$W_\emptyset$ in terms of gluon scattering amplitudes is possible,
as was shown by Alday and Maldacena in Ref.~\cite{Alday:2007he}.
There is no sense to the concept of scattering amplitudes in a
conformal field theory such as ${\cal
  N}=4$; conformal invariance precludes the existence of asymptotic
states. The reflection of this issue is the appearance of IR
divergences corresponding to the exchange of soft particles between
the external particle legs. The introduction of an IR regulator
breaks conformal invariance and allows the definition and
computation of these scattering amplitudes. This can be achieved in
the string dual of ${\cal N}=4$ SYM through the introduction of a
D3-brane at a small distance $r_0$ from the center of $AdS_5$,
effectively introducing into the spectrum a gap or IR cut-off
proportional to $r_0$. Upon such a D3-brane one can consider the
scattering of open strings. In fact, due to the warping of the $AdS$
metric these are strings of large proper momentum, and as for the
case of flat space, their scattering is captured by the growth of a
macroscopic connected string worldsheet joining the incoming and
outgoing strings. The picture is completed through a T-duality
transformation on the boundary coordinates which inverts the $AdS_5$
space into another $AdS_5$, the IR regulator of the former becoming
a UV regulator of the latter - the familiar cut-off associated with
the approach to the boundary in AdS/CFT. What one has in the new
$AdS_5$ is then a macroscopic worldsheet which ends upon a contour
at the boundary. This contour is defined by the momenta $k^\m_i$ of
the incoming and outgoing strings in the original $AdS$. Under the
action of the T-duality, they are mapped to points $y^\m_i$ on the
boundary of the new $AdS$ through
\be y_{i+1} - y_i = 2\pi k_i \ee
The shape of the resulting Wilson loop is then evident: the
conservation of the $k^\m$ implies a closed contour, the null-ness
of the $k^\m$ (i.e. $k^2=0$) implies light-like segments.

\begin{figure}
\begin{center}
\includegraphics*[bb= 0 0 320 155,height=1.25in]{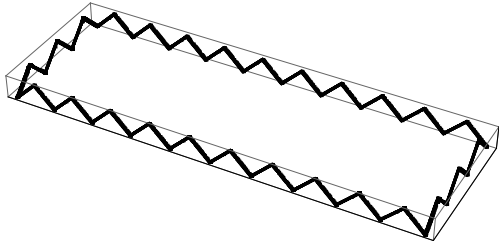}
\end{center}
\caption{An example of a null zig-zag Wilson loop. Here the smooth
  underlying curve approximated by the zig-zag is a spacelike rectangle.}
\label{fig-rect}
\end{figure}

\begin{figure}
\begin{center}
\includegraphics*[bb= 0 18 111 140,height=1.25in]{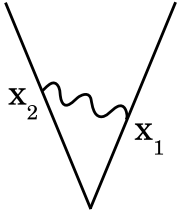}
\end{center}
\caption{Diagram giving rise to a cusp divergence at 1-loop. The two
  edges are null segments of a null zig-zag Wilson loop.}
\label{fig-cusp}
\end{figure}

In \cite{Alday:2007he} Alday and Maldacena considered a null zig-zag
which approximated a smooth (spacelike) rectangle, see
Fig.~\ref{fig-rect}. They argued that in the limit of large $n$,
where $n$ counts the number of null segments, the finite part of the
vacuum expectation value should match that of a Wilson loop without
coupling to scalars, whose contour is the underlying smooth
space-like path.
For the rectangular Wilson loop, they observed that one can see this
explicitly at leading order in perturbation theory. We will show this
example presently, as it serves as a warm-up for the circle which we
will consider afterwards. Before we do this however we must qualify
what is meant by the ``finite part'' of the expectation value. The
divergences in the gluon scattering amplitudes mentioned above
manifest themselves in the null zig-zag as cusp divergences. At
leading order in perturbation theory one can see these cusp
divergences by considering the gluon exchange between two null
segments which meet at an angle, see Fig.~\ref{fig-cusp}.
We will now evaluate this diagram. Using the Feynman gauge and
dimensional regularization, the gluon propagator is
\be \la A_\m^a(x) \, A_\n^b(0)\ra = -\frac{\G(1-\e)}{4\pi^{2-\e}}
\frac{g^2
\m^{2\e}\,\d^{ab}\,\eta_{\m\n}}{[-x^2+i\varepsilon]^{1-\e}} \ee
where $\eta_{\m\n} = (+,-,-,-)$ and the dimension of spacetime is
$4-2\e$. Expanding (\ref{Wnull}) to second order and taking its
expectation value, one finds for the diagram shown in figure
\ref{fig-cusp} the following contribution
\bsp\label{cusp}
\la W_{\emptyset} \ra &= 1 + \frac{\l(\pi \m^2)^{\e}}{4}\frac{\G(1-\e)}{4\pi^2}\int_0^1 dt_1\int_0^1 dt_2\,\frac{\dot x_1 \cdot \dot x_2}{[-(x_1-x_2)^2]^{1-\e}}\\
&= 1 - \l(\pi \m^2)^{\e} \frac{\G(1-\e)}{16\pi^2}\int_0^1 dt_1\int_0^1 dt_2\,\frac{(k_1 \cdot k_2)^\e}{[2t_1(1-t_2)]^{1-\e}}\\
&=1 -\frac{\l}{16\pi^2} \frac{(\m^2\pi s)^\e \,\G(1-\e)}{2\e^2}
\end{split}
\ee
where $x_1^\m = t_1 \, k_1^\m$, $x_2^\m = (1-t_2)\,k_2^\m$, and $s =
(k_1+k_2)^2$; the $k_i$ being null momenta. Clearly by expanding in
$\e$ we will also produce finite terms of order $\e^0$. When drawing
the correspondence between $\la W_{\emptyset} \ra$ and the
non-scalar-coupled Wilson loop, we ignore these finite terms. We are
interested in those finite pieces which do not scale with the number
of cusps $\nz$. At one-loop we therefore consider only the non-cusp
contributions to the expectation value. These are finite, and in the
limit of large $\nz$ will produce both $\nz$-dependent and
$\nz$-independent terms
 \be
\lim_{\nz\to\infty} \Bigl[ \la W_{\emptyset} \ra^{1-\text{loop}}
- \la W_{\emptyset} \ra^{1-\text{loop}}_{\text{cusps}} \Bigr] = A\,
\nz^2 +B\, \nz + C + {\cal O}(\nz^{-1}).
 \ee
The non-scalar-coupled Wilson loop is also divergent for spacelike
contours. Indeed, the scalar coupling in Eq.~(\ref{W}) may be viewed
as a regularization of the non-scalar-coupled loop: the loop-to-loop
propagator in Eq.~(\ref{W}) is proportional to $(|\dot x_1||\dot
x_2| - \dot x_1 \cdot \dot x_2)/(x_1-x_2)^2$, for smooth contours,
this object has no singularity when $x_1 \to x_2$. We will compute
the ``finite-part'' of the non-scalar-coupled Wilson loop by
dimensional regularization. The statement which we shall prove is
then
 \be\label{state}
 C = \la W_{NSC} \ra^{1-\text{loop}}_{\text{finite}}
  \ee
where $W_{NSC}$ is the non-scalar-coupled Wilson loop whose smooth,
spacelike contour is approximated by the null zig-zag.

\subsection{Summary of results}
We summarize the results we obtain here, and leave the detailed
calculations to the following section.
\begin{itemize}
\item{Anti-parallel lines}\\
 We find, by explicit calculation, the relation Eq.~(\ref{state}) true for the contour being two anti-parallel
 lines.
\item{Circular loop}\\
 We calculate the one-loop expectation value of \w on null zigzags approximating
 a circle, again confirming the relation Eq.~(\ref{state}).
\item{Non-parallel lines}\\
 We consider the contour composed of two separated space-like line
 segments at an arbitrary angle. We find that the
 relation Eq.~(\ref{state}) is indeed correct. We argue that this
 result allows one to extend Alday \& Maldacena's argument
 \cite{Alday:2007he} to any smooth, space-like shape.
\end{itemize}

\subsection{Outline}
The rest of the paper is organized as follows. In section 2, we review
the equality, ref Eq.~(\ref{state}) for anti-parallel lines. In
section 3, we consider two separated lines at an arbitrary angle. In
section 4, we consider a circular contour.  With these relations, we
argue for the extension of the equality to any smooth, space-like
shape in section 5.
\section{Anti-parallel lines}

As discussed in the previous section the simplest setting in which
to prove our result is the anti-parallel lines; indeed this
case was already considered in \cite{Alday:2007he}. We present the
calculation in some detail as it serves to clarify the other
contours considered in subsequent sections.

First we specify coordinates of the anti-parallel lines and their \z
approximation. One straight line spans the two points $(0,0,0,0)$
and $(0,L,0,0)$. Its \z sequence consists of points
\be\label{seq}
\left\{x_{(1)},y_{(1)},z_{(1)},x_{(2)},y_{(2)},z_{(2)},\ldots ,
x_{(\nz)},y_{(\nz)},z_{(\nz)}\right\} \ee
where $\nz$ is some integer and $\nz \gg 1$ and
\be x_{(j)}^\m = \frac{L}{\nz}\Bigl( 0, (j-1),0,0 \Bigr),~~
y_{(j)}^\m = \frac{L}{\nz}\Bigl( 1/2,(j-1/2),0,0 \Bigr),~~
z_{(j)}^\m = \frac{L}{\nz} \Bigl( 0,j,0,0\Bigr) \ee
specifies the null zig-zag approximation as $y_{(j)} - x_{(j)}$ and
$z_{(j)} - y_{(j)}$ are null vectors. We may then specify another
straight line spanning $(0,0,T,0)$ and $(0,L,T,0)$ with barred
variables $\bar x_{(j)}$, $\bar y_{(j)}$, $\bar z_{(j)}$, defined as
$\bar x_{(j)} = x_{(j)} + (0,0,T,0)$ and similarly for $\bar
y_{(j)}$ and $\bar z_{(j)}$.

\begin{figure}
\begin{center}
\includegraphics*[bb= 31 41 408 324,height=1.25in]{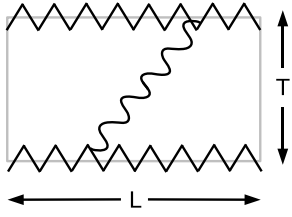}
\end{center}
\caption{Finite contribution to the null zig-zag rectangle at
1-loop.} \label{fig-rectdiag}
\end{figure}

We begin by considering those contributions stemming from the
diagram pictured in Fig.~\ref{fig-rectdiag}. On the bottom straight
line the gluon line is attached to the segment $x_{(i)} +(y_{(i)} -
x_{(i)}) \,t$, while on tope one it is attached to $\bar z_{(j)}
+(\bar y_{(j)}- \bar z_{(j)}) \,\bar t$.

We find the following result\footnote{We have set $\e$ to zero here
  since the diagram is finite.}
\bsp
\la W \ra_{x_{(i)}\,\bar y_{(j)}} 
&= \frac{\l}{16\pi^2} \int_0^1 dt \int_0^1 d\bar t
\,\frac{1}{2}\,\frac{1}{\bigl(i-j - t\bigr)\bigl(i-j -\bar{t}\bigr)
+ T^2  \nz^2/L^2}
\end{split}
\ee
The contribution from $\la W \ra_{y_{(i)}\,\bar x_{(j)}}$ will yield
the same result, once we have summed over $i$ and $j$. The
contributions $\la W \ra_{x_{(i)}\,\bar x_{(j)}}$ and $\la W
\ra_{y_{(i)}\,\bar y_{(j)}}$ are zero, as the null-momenta defining
the two null segments are (anti-)parallel in this case and thus have
zero inner product. We will show that in the large $\nz$ limit the
contributions
\be\label{rectangletwo} \sum_{i=1}^{\nz} \sum_{j=1}^{\nz} \,\left(
\la W \ra_{x_{(i)}\,\bar
  y_{(j)}}+ \la W \ra_{y_{(i)}\,\bar x_{(j)}}\right)=
2 \sum_{i=1}^{\nz} \sum_{j=1}^{\nz} \, \la W \ra_{x_{(i)}\,\bar
y_{(j)}} \ee
are equal to the analogous diagram in the computation of $\la
W_{NSC}\ra$. In the large $\nz$ limit Eq.~(\ref{rectangletwo})
becomes
\bsp\label{rectresult}
&\lim_{\nz\to\infty}\frac{\l}{16\pi^2}\sum_{i=1}^{\nz} \sum_{j=1}^{\nz} \,\int_0^1 dt \int_0^1 d\bar t \frac{1}{\bigl(i-j - t\bigr)\bigl(i-j -\bar{t}\bigr) + T^2  \nz^2/L^2}\\
&=\!\frac{\l}{16\pi^2}\! \lim_{\nz\to\infty}\!\sum_{k=1}^{\nz}\!\!\int_0^1\!\!\!\int_0^1\!dtd\bar t\left(\frac{\nz\!-\!k}{(k\!+\!t)(k\!+\!\bar{t})\!+\!T^2\!\nz^2\!/\!L^2}\!+\big(\!t\to -t,\bar{t}\to -\bar{t}\!\big)+\frac{\nz}{t\bar{t}\!+\!T^2\!\nz^2\!/\!L^2}\right)\\
&=\frac{\l}{16\pi^2}\!\!\int_0^1\!\!\int_0^1\!dtd\bar t\lim_{\nz\to\infty}\frac{1}{\sqrt{g^2-4h}}\bigg\{f(g/2+\sqrt{g^2/4-h})\!-f(-g/2\!-\!\sqrt{g^2/4-h})\\
&~~~~~~~~~~~~~~+f(-\!g/2+\sqrt{g^2/4-h})-f(g/2-\sqrt{g^2/4-h})\bigg\}-\frac{\l}{16\pi^2}\nz\rm{Li_2}(-\frac{T^2}{\nz^2L^2})\\
&=\frac{\l}{16\pi^2}\, \bigg(
2\frac{L}{T}\arctan\frac{L}{T}-\log\big(1+\frac{L^2}{T^2}\big)
\bigg)+ {\cal O}(\nz^{-1})
\end{split}
\ee
where $g\equiv(t+\bar{t})$, $h\equiv(t\bar{t}+T^2 \nz^2/L^2)$,
$f(x)\equiv(x+\nz)(\psi(1+x)-\psi(1+x+\nz))$, $\psi$ is $Digamma$
function; $\rm{Li_2}$ is $Dilogarithm$ function.
In third line of Eq.~(\ref{rectresult}) we have performed the
summation over $k$ analytically, then in the final line taken large
$\nz$ limit.  This expression is precisely the contribution stemming
from the diagram in Fig.~\ref{fig-rectdiag}, however with a
space-like anti-parallel lines (shown in gray) replacing the null
zig-zag, i.e.,
\bsp\label{NSCrect} \la W_{NSC}\ra &\to \frac{\l}{16\pi^2} \int_0^1d t \int_0^1 d\bar t\,\frac{L^2}{L^2(t-\bar t)^2 + T^2 }\\
&=\frac{\l}{16\pi^2}\left(
2\frac{L}{T}\arctan\frac{L}{T}-\log\big(1+\frac{L^2}{T^2}\big)
\right).
\end{split}
\ee
%
%
\section{Non-parallel lines}
\label{sec-ArbCont}
\begin{figure}
\begin{minipage}[t]{0.5\linewidth}
\centering
\includegraphics[width=4cm,clip=true,keepaspectratio=true]{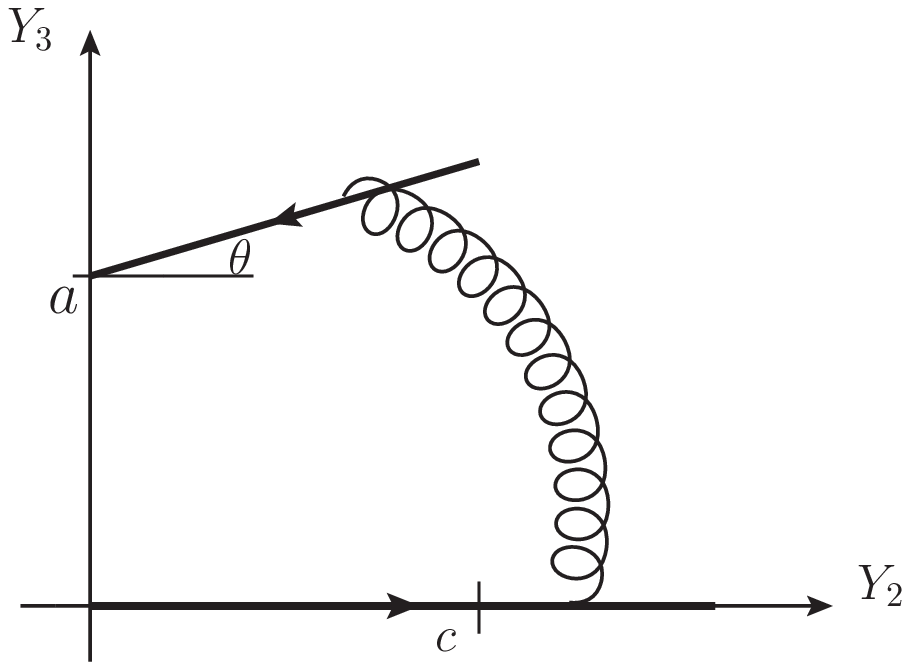}
 \caption{  Two space-like segments \newline at an angle
$\th$.}\label{fig:twoline}
\end{minipage}%
\begin{minipage}[t]{0.5\linewidth}
\centering
\includegraphics[width=5cm,clip=true,keepaspectratio=true]{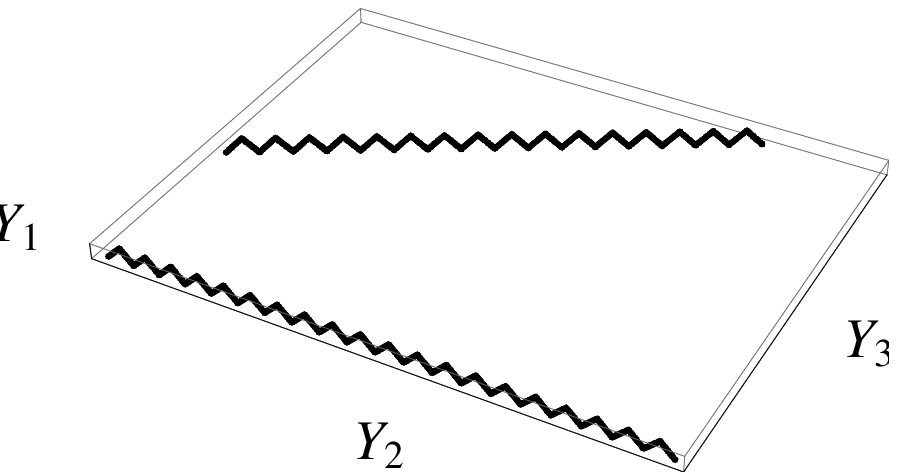}
\caption{Fig.~\ref{fig:twoline} is approximated by null \z
sements.}\label{fig:twolineapproximation}
\end{minipage}
\end{figure}

 In order to generalize Alday \& Maldacena's argument to any smooth,
 space-like curve, we should consider two separated segments
 at one angle $\theta$, see Fig.~\ref{fig:twoline}. In this section we will compare the finite parts
 of the gluon exchange between the two separated segments to that of
 the null \z approximation shown in Fig.~\ref{fig:twolineapproximation}.
The length of the horizontal segment is taken to be $1$, while the
distance between the two segments along the vertical axis is given
by $a$. We parameterize their \z approximations as follows
 \be\label{twolinecoordinateappro}
 \begin{split}
  x_{(i)}&=\frac{1}{\nz}\left(0,i-1,0,0\right),~y_{(i)}=\frac{1}{\nz}\left(1/2,i-1/2,0,0\right),~z_{(i)}=\frac{1}{\nz}\left(0,i,0,0\right),\\
 \bar x_{(i)}&=\frac{1}{\nz}\left(0,(i-1),(i-1)\tan\th +a \nz,0\right),~~\bar z_{(i)}=\frac{1}{\nz}\left(0,i,i\tan\th +a \nz,0\right),\\
  \bar y_{(i)}&=\frac{1}{\nz}\left(1/2\sec\th,(i-1/2),(i-1/2)\tan\th + a \nz, 0\right)
 \end{split}
  \ee
  and take the index on barred variables to run to $cN$, where $c$ is
  some rational number accounting for the different length
  of the angled segment.  The one-loop expectation value of \w in the
  large $\nz$ limit is given by
\be
 \la W_{\emptyset}\ra=\frac{\l}{16\pi^2}\,\lim_{\nz\to\infty}\sum_{i=1}^{\nz}\sum_{j=1}^{c\nz} \bigg(\la W\ra_{y_{(i)}\,\bar x_{(j)}}+\la W\ra_{x_{(i)}\,\bar y_{(j)}}+\la W\ra_{x_{(i)}\,\bar x_{(j)}}+\la W\ra_{y_{(i)}\,\bar y_{(j)}}\bigg)\\
\ee
with
 \bsp\label{twoline}
&\la W \ra_{y_{(i)}\bar x_{(j)}}\!=\!\int_0^1\!\!\!\int_0^1\!\!dt\,d\!s\,\frac{1+\sec\th}{(2i\!-\!2j\!+\!2\!-\!s\!-\!t)^2\!+\!(2a\nz\!+\!(2j\!-\!2\!+\!t)\tan\th)^2\!-\!(s\!-\!t\sec\th)^2}\\
&\la W \ra_{x_{(i)}\,\bar y_{(j)}}\!=\!\int_0^1\!\!\!\int_0^1\!\!dt\,d\!s\, \frac{1+\sec\th}{(2i\!-\!2j\!-\!2\!+\!s\!+\!t)^2\!+\!(2a\nz\!+\!(2j\!-\!t)\tan\th)^2\!-\!(s\!-\!t\sec\th)^2}\\
&\la W \ra_{x_{(i)}\,\bar x_{(j)}}\!=\!\int_0^1\!\!\!\int_0^1\!\!dt\,d\!s\, \frac{1-\sec\th}{(2i\!-\!2j\!+\!s\!-\!t)^2\!+\!(2a\nz\!+\!(2j\!-\!2\!+\!t)\tan\th)^2\!-\!(s\!-\!t\sec\th)^2}\\
&\la W \ra_{y_{(i)}\,\bar
y_{(j)}}\!=\!\int_0^1\!\!\!\int_0^1\!\!dt\,d\!s\,
\frac{1-\sec\th}{(2i\!-\!2j\!-\!s\!+\!t)^2\!+\!(2a\nz\!+\!(2j\!-\!t)\tan\th)^2\!-\!(s\!-\!t\sec\th)^2}
 \end{split}
 \ee

Note that the last term in each of the denominators in
Eq.~(\ref{twoline}) is small compared to the remaining terms; it can
be used to expand these integrands. One can check that the
correction terms are order of $\nz^{-4}$, thus their contributions
to $\la W_{\emptyset}\ra$ are negligible in the large $\nz$ limit.
Then we have
 \bsp\label{twolines}
&\la W \ra_{y_{(i)}\bar x_{(j)}}\!=\!\int_0^1\!\!\!\int_0^1\!\!dt\,d\!s\,\frac{1+\sec\th}{(2i\!-\!2j\!+\!2\!-\!s\!-\!t)^2\!+\!(2a\nz\!+\!(2j\!-\!2\!+\!t)\tan\th)^2}+\mathcal{O}(\nz^{-4})\\
&\la W \ra_{x_{(i)}\,\bar y_{(j)}}\!=\!\int_0^1\!\!\!\int_0^1\!\!dt\,d\!s\, \frac{1+\sec\th}{(2i\!-\!2j\!-\!2\!+\!s\!+\!t)^2\!+\!(2a\nz\!+\!(2j\!-\!t)\tan\th)^2\!}+\mathcal{O}(\nz^{-4})\\
&\la W \ra_{x_{(i)}\,\bar x_{(j)}}\!=\!\int_0^1\!\!\!\int_0^1\!\!dt\,d\!s\, \frac{1-\sec\th}{(2i\!-\!2j\!+\!s\!-\!t)^2\!+\!(2a\nz\!+\!(2j\!-\!2\!+\!t)\tan\th)^2\!}+\mathcal{O}(\nz^{-4})\\
&\la W \ra_{y_{(i)}\,\bar
y_{(j)}}\!=\!\int_0^1\!\!\!\int_0^1\!\!dt\,d\!s\,
\frac{1-\sec\th}{(2i\!-\!2j\!-\!s\!+\!t)^2\!+\!(2a\nz\!+\!(2j\!-\!t)\tan\th)^2\!}+\mathcal{O}(\nz^{-4})
 \end{split}
 \ee
Next we intend to prove that integrations in Eq.~(\ref{twolines})
can be simplified to a desired form under summation over $i,j$, for
example
\bsp\label{arbitraryappro}
 &\sum_{j=1}^{c\nz}\sum_{i=1}^\nz\!\int_0^1\!\!\!\int_0^1\!\!dt\,d\!s\,\frac1{(2i\!-\!2j\!+\!2\!-\!s\!-\!t)^2\!+\!(2a\nz\!+\!(2j\!-\!2\!+\!t)\tan\th)^2}\\
 &~~~~~~=\sum_{j=1}^{c\nz}\sum_{i=1}^\nz \frac{1/4}{(i\!-\!j)^2\!+\!(a\nz\!+\!j\tan\th)^2}+\mathcal {O}(\nz^{-1})
 \end{split}
 \ee

Suppose $c>1$, the left-hand side of Eq.~(\ref{arbitraryappro}) is
 \bsp
 &\bigg\{\sum_{j=1}^{\nz-1}\sum_{i=j+1}^\nz+\sum_{j=1}^\nz\sum_{i=1}^{j-1}+\sum_{j=\nz+1}^{c\nz}\sum_{i=1}^{\nz}+\sum_{i=j=1}^\nz\bigg\}\int_0^1\!\!\!\int_0^1\!\!dt\,d\!s\,\\
 &~~~\frac1{(2i\!-\!2j\!+\!2\!-\!s\!-\!t)^2\!+\!(2a\nz\!+\!(2j\!-\!2\!+\!t)\tan\th)^2}\equiv \alpha+\beta+\delta+\zeta
 \end{split}
 \ee
From the definition of $\alpha$, we find
 \bsp\label{ab}
 \alpha&>\sum_{j=1}^{\nz-1}\sum_{i=j+1}^\nz \frac{1}{(2i-2j+2)^2+(2a\nz+2j\tan\th)^2}\\
&=\bigg\{\sum_{j=1}^{\nz-1}\sum_{i=j+1}^\nz-\sum_{j=1}^{\nz-1}(i=j+1)+\sum_{j=1}^{\nz-1}(i=\nz+1)\bigg\}\frac{1/4}{(i-j)^2+(a\nz+j\tan\th)^2}\\
&=\sum_{j=1}^{\nz-1}\sum_{i=j+1}^\nz\frac{1/4}{(i-j)^2+(a\nz+j\tan\th)^2}+\mathcal{O}(\nz^{-1})
  \end{split}
 \ee
and
 \bsp\label{as}
\alpha&<\sum_{j=1}^{\nz-1}\sum_{i=j+1}^\nz\frac{1/4}{(i-j)^2+(a\nz+(j-1)\tan\th)^2}\\
&=\bigg\{\sum_{j=1}^{\nz-1}\sum_{i=j+1}^\nz-\sum_{j=1}^{\nz-1}(i=\nz)+\sum_{i=1}^{\nz-1}(j=0)\bigg\}\frac{1/4}{(i-j)^2+(a\nz+j\tan\th)^2}\\
&=\sum_{j=1}^{\nz-1}\sum_{i=j+1}^\nz\frac{1/4}{(i-j)^2+(a\nz+j\tan\th)^2}+\mathcal{O}(\nz^{-1})
 \end{split}
 \ee
So in the large $\nz$ limit we have
 \be\label{afinal}
 \alpha=\sum_{j=1}^{\nz-1}\sum_{i=j+1}^\nz\frac{1/4}{(i-j)^2+(a\nz+j\tan\th)^2}+\mathcal{O}(\nz^{-1})
 \ee
This approach to simplify $\alpha$ as Eq.~(\ref{afinal}) works well
for $\beta,~\delta,~\zeta$. Putting these simplified expressions
together, we get the right-hand side of Eq.~(\ref{arbitraryappro}).
If $c\leqslant 1$, the double summation over $i,j$ can be grouped in
a different way, but one can repeat the process as
Eq.~(\ref{ab}-\ref{afinal}) to verify Eq.~(\ref{arbitraryappro}).

Using the same method as in the other expressions in Eq.~(\ref{twolines}),
we obtain
 \bsp
 \la W_{\emptyset} \ra&=
\frac{\l}{16\pi^2}\sum_{i=1}^{\nz}\sum_{j=1}^{c
  \nz}\frac{1}{(i-j)^2+(a\nz+j\tan\th)^2}
+\mathcal {O}(\nz^{-1})\\
&\to\frac{\l}{16\pi^2}\int_0^1dt \int_0^c
ds\frac{1}{(s-t)^2+(a+s\tan\th)^2}
 \end{split}
 \ee
This result is just the expression when one calculate one-loop
expectation value of \w  defined on Fig.~\ref{fig:twoline}.
\section{Circle}\label{sec:circle}

\begin{figure}
\begin{center}
\includegraphics*[bb= 0 0 320 210,height=1.25in]{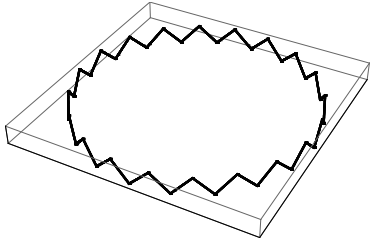}
\end{center}
\caption{Null zig-zag Wilson loop which approximates a spacelike
circle.} \label{fig-circle}
\end{figure}

We approximate a space-like circle of unit radius with a null
zig-zag specified by a sequence points as in Eq.~(\ref{seq}) with
\bsp
x_{(j)}^\m &= \Bigl( 0, 0,-\cos(j+1/2)\vp,\sin(j+1/2)\vp \Bigr);\\
y_{(j)}^\m &= \Bigl( \sin (\vp/2),0,-\cos(\vp/2)\cos j\vp,\cos(\vp/2)\sin j\vp \Bigr);\\
z_{(j)}^\m &= \Bigl( 0, 0,-\cos(j-1/2)\vp,\sin(j-1/2)\vp \Bigr)
\end{split}
\ee
where $ \vp=2\pi/\nz$ since the contour is closed, see
Fig.~\ref{fig-circle}.
The one-loop expectation value of the null zig-zag \w in the large
$\nz$ limit is given by
\be\label{circl}
 \la W_{\emptyset}\ra^{1-\text{loop}} =\frac{\l}{16\pi^2}\lim_{\nz\to\infty} \sum_{i,j=1}^\nz \Bigl(\la W \ra_{x_{(i)}\, x_{(j)}} +\la W \ra_{y_{(i)}\, y_{(j)}}+\la W \ra_{x_{(i)}\, y_{(j)}} +\la W \ra_{y_{(i)}\, x_{(j)}}\Bigr)\\
 \ee
where $\la W \ra_{x_{(i)}\, y_{(j)}}$ stands for the contribution
from $x_{(i)}+t(y_{(i)}-x_{(i)})$ and $y_{(j)}+t(z_{(j)}-y_{(j)})$,
$\la W \ra_{x_{(i)}\, x_{(j)}}=\la W \ra_{y_{(j)}\, y_{(i)}}$ and
$\la W \ra_{x_{(i)}\, y_{(j)}}=\la W \ra_{y_{(i)}\, x_{(j)}}$.
\bsp\label{circlzz1}
\la\!W\!\ra_{x_{(i)}x_{(j)}}\!=\!\int_0^1\!\!\!\int_0^1\!\!dt\,d\!s\frac{-\sin((i\!-\!j)\vp/\!2)\eta^2/\!2}{(s\!-\!t)\eta\cos((i\!-\!j)\vp/\!2)\!-\!(1\!+\!s t \eta^2 )\sin((i\!-\!j)\vp/\!2)}\\
\end{split}
 \ee
\bsp\label{circlzz2}
\la\!W\!\ra_{x_{(i)}y_{(j)}}\!=\!\int_0^1\!\!\!\int_0^1\!\!dt\,d\!s\frac{-\cos((i\!-\!j)\vp/\!2)^2\!\eta^2/\!2}{\sin((\!i-\!j)\!\vp/2)^2\!+\!st\eta^2\!\cos((i\!-\!j)\vp/2)^2\!-\!\eta(\!s\!+\!t)\!\sin((\!i\!-\!j)\vp)/\!2}\\
\end{split}
 \ee
where $\eta\equiv-\tan(\vp/2)$.  The integration over $s,t$ can be
preformed straightforward, but it isn't easy to obtain analytic
result by summing over $i, j$ from that. We will try another
approach.

 Let us first consider the contribution from $\la W \ra_{x_{(i)}\,
 x_{(j)}}$, notice that there is no cusp arising. From
 Eq.~(\ref{circlzz1}), we have\footnote{In second line, we have calculated $k=1$ case individually. The result is $\nz$
 $\{\log2(\log(1+a)-\log(1-a))
 +\log(1+a)\log(1-a)+\rm{Li_2}(a)+\rm{Li_2}(a/(1+a))-\rm{Li_2}(2a/(1+a))\}$, here $a\equiv\tan^2(\pi/N)$. Its contributions is vanishing in the large $\nz$ limit.}
\bsp\label{circlzz1s}
 \sum_{i,j=1}^{\nz} \la W \ra_{x_{(i)}\, x_{(j)}}\!&=\sum_{k=1}^{\nz/2-1}\!\int_0^1\!\!\!\int_0^1\!\!dt\,d\!s\frac{ \nz\tan(\pi/\nz)^2}{(s-t)b+1+st\tan(\pi/\nz)^2 }\\
 &=\sum_{k=2}^{\nz/2-1}\!\int_0^1\!\!\!\int_0^1\!\!dt\,d\!s\frac{ \nz\tan(\pi/\nz)^2}{(s-t)b+1+st\tan(\pi/\nz)^2 }+\mathcal  {O}(\nz^{-1})\\
&=\nz\tan^2\frac{\pi}{\nz}\!\sum_{k=2}^{\nz/2-1}\!\int_0^1\!\!\!\int_0^1\!\!dt\,d\!s\frac1{(s-t)b+1 }+\mathcal  {O}(\nz^{-1})\\
&=\nz\tan^2\frac{\pi}{\nz}\!\sum_{k=2}^{\nz/2-1}\frac{(1+b)\log(1+b)+(1-b)\log(1-b)}{b^2}\\
&=\nz\tan^2\frac{\pi}{\nz}\sum_{z=1}^\infty\left(\frac{2}{2z-1}-\frac1{z}\right)\sum_{k=2}^{\nz/2-1}b^{2z-2}
 \end{split}
 \ee
where $b\equiv\tan(\pi/\nz)\cot(k\pi/\nz)$. When $k\in[2,\nz/2-1]$,
$(s-t)b+1\in[1/2,3/2]$, thus we can use $st\tan(\pi/\nz)^2$ as a
parameter to expand the integrand in second line and find the
correction terms are vanishing in the large $\nz$ limit. In
addition, $b\in[0,1/2]$ can be used to expand the functions
$\log(1\pm b)$ as indicated in last equation of
Eq.~(\ref{circlzz1s}).

 For any integer $z\geqslant1$, analytic result of
 $\sum_{k=2}^{\nz/2-1}b^{2z-2}$ can be obtained. We find that only $z=1$ will
 survive in the large $\nz$ limit. Thus we have,
\bsp
 \lim_{\nz\to\infty}\sum_{i,j=1}^{\nz} \la W \ra_{x_{(i)}\, x_{(j)}}=\frac{\pi^2}{2}+\mathcal  {O}(\nz^{-1})
 \end{split}
 \ee

Now consider $\la W \ra_{x_{(i)}\, y_{(j)}}$, we need take out the
contribution from cusp divergence when $i=j,j-1$. From
Eq.~(\ref{circlzz2}), we get
 \bsp
&\sum_{i\neq j,j-1}^{\nz}\!\!\!\la W \ra_{x_{(i)}\,y_{(j)}}=\frac{\nz}{2}\sum_{\substack{k=1\\k\neq N/2}}^{\nz-2}\int_0^1\!\!\int_0^1dt\,d\!s \frac1{s-t}\left(\frac{1}{\cot(\pi/\nz)\tan(k\pi/\nz)+s}-(s\to t)\right)\\
&~~~=2\nz\sum_{k=2}^{\nz/2-1}\int_0^1 d\!s\log\frac{1-s}{s}\sum_{z=1}^\infty\frac{s^{2z-1}}{(\cot(\pi/\nz)\tan(k\pi/\nz))^{2z}}-\frac{\nz}{2}\log^22\\
&~~~=\nz\sum_{z=1}^\infty
\frac{1-2z\,H(2z)}{4z^2}\bigg\{\tan^{2z}(\pi/\nz)\sum_{k=1}^{\nz-1}\cot^{2z}(k\pi/\nz)-2\bigg\}-\frac{\nz}{2}\log^22
\end{split}
 \ee
where $H(x)\equiv\psi(1+x)+\gamma$, $\psi$ is $Digamma$ function and
$\gamma$ is Euler constant.

 We find in the large $\nz$ limit,
 \bsp
\lim_{\nz\to\infty}\sum_{i\neq j,j-1}^{\nz} \la W \ra_{x_{(i)}\,
y_{(j)}}=\#\nz+\frac{\pi^2}{2}+\mathcal{O}(\nz^{-1})
\end{split}
 \ee
 with $\#$ being a real finite number,
 \bsp
  \#\equiv\sum_{z=1}^\infty \frac{1-2z\,H(2z)}{4z^2}\left\{(-1)^{z+1}2^{2z}\pi^{2z}\frac{B_{2z}}{(2z)!}-2\right\}-\frac1{2}\log^22
\end{split}
 \ee
 where $B_n$ is the nth Bernoulli number.
  After these lengthy algebra, we obtain
 \bsp
 &\Bigl[ \la W_{\emptyset} \ra^{1-\text{loop}}-\la W_{\emptyset} \ra^{1-\text{loop}}_{\text{cusps}} \Bigr]=\frac{\l}{16\pi^2}2\left(\pi^2+\#\nz\right)+\mathcal {O}(\nz^{-1})~\sim\frac{\l}{8}\\
 \end{split}
 \ee
 up to some term proportional to $\nz$. We also can reproduce this finite part for expectation value of \w
 defined on a space-like circle. Parameterize a circle with unit radius by $x^\mu(t)=(0, \cos2\pi t,
 \sin2\pi t,0)$. By definition, we find its one-loop expectation
 value,
 \bsp
\la W_C \ra_{NSC}&=\frac{\l(\pi \m^2)^{\e}}{4}\frac{\G(1-\e)}{4\pi^2}\int_0^1 d t \int_0^1 d s\,\frac{\dot{x}(t)\cdot\dot{x}(s)}{[-(x(t)-x(s))^2]^{1-\e}}\\
&=-\frac{\l(\pi \m^2)^{\e}}{4}\frac{\G(1-\e)}{4\pi^2}\frac{2\pi}{2^{1-\e}}\int_0^{2\pi} d \th \,\left\{\sum_{k=0}^{\infty}\frac1{k!}(\cos\th)^{k+1}\frac{\Gamma(1+k-\e)}{\Gamma(1-\e)}\right\}\\
&=-\frac{\l(\pi\m^2)^{\e}}{4}\frac{\G(1-\e)}{4\pi^2}\frac{2\pi}{2^{1-\e}}\,\frac{2^\e\sqrt{\pi}\Gamma(2-\e)\Gamma(-1/2+\e)}{\Gamma(1-\e)\Gamma(1+\e)}=1+\frac{\l}{8}+\mathcal
{O}(\e^1)
 \end{split}
 \ee
%
\section{Comment on generalization to arbitrary curves}

We have verified the Eq.~(\ref{state}) on two separated lines in
section \ref{sec-ArbCont}. This result should be able to be used to
generalize this equation to any smooth, space-like shape as follows.
First one can approximate a smooth space-like curve by a spacelike
$n$-sided polygon. Building upon that polygon a zig-zag approximation,
we know from the results of section \ref{sec-ArbCont} that the finite
part of the non-scalar-coupled space-like Wilson loop is recovered.
Then taking $n$ to scale like the number of zig-zags, any effects
associated with the corners of the $n$-gon, or self-interactions of
the $n$-gon edges themselves will contribute only to the piece which
scales as the number of zig-zags. The finite part should then
reproduce the finite part of the smooth, non-scalar-coupled,
space-like curve being approximated.  Indeed, we have verified that
this proposal for generalizing to general curves works for the case of
a circle, by using a regular $n$-gon as an approximation and applying
the results of section \ref{sec-ArbCont}.

\section*{Acknowledgements}

I am grateful to Jan Plefka, Donovan Young, and the rest of the QFT
group at Humboldt University, Berlin where this work was started. I
would also like to thank Prof.  Wu Yue-liang,  Wu Feng, Yang Yibo
for helpful discussions. This work was supported by the Chinese
Academy of Sciences and the Deutscher Akademischer Austausch Dienst
(DAAD).


\end{document}